\documentclass[preprint2]{aastex}
\usepackage{amssymb, amsmath, graphics} 
\usepackage{graphicx} 
\usepackage{aas_macros} 
 
\usepackage[usenames]{color}

\def\Teff{$T_{\mathrm{eff}}$}
\def\logg{\ensuremath{\log g}}
\def\vmic{$\upsilon_{\mathrm{mic}}$}

\def\vsini{\ensuremath{{\upsilon}\sin i}}
\def\kms{$\mathrm{km\,s}^{-1}$}

\def\vr{${\upsilon}_{\mathrm{r}}$}

\def\llm{{\sc LLmodels}}

\def\synth{{\sc SYNTH3}}

\def\errTeff{$\sigma_{T_{\mathrm{eff}}}$}

\shorttitle{Hybrid pulsator KIC\,8054146}
\shortauthors{Michel Breger et al.}

\begin{document}

\title{Relationship between low and high frequencies in Delta Scuti stars: Photometric {\it Kepler} and spectroscopic analyses of the rapid rotator KIC\,8054146.\thanks{Based on observations 
obtained with the Hobby-Eberly Telescope, which is a joint project of the University of Texas at Austin, the Pennsylvania State University, Stanford University, Ludwig-Maximilians-Universit\"at M\"unchen, and Georg-August-Universit\"at G\"ottingen.}} 
 
\author{M.~Breger\altaffilmark{1,2}, L. Fossati\altaffilmark{3}, L.~Balona\altaffilmark{4}, D. W. Kurtz\altaffilmark{5},  P. Robertson\altaffilmark{1}, D. Bohlender\altaffilmark{6},
P.~Lenz\altaffilmark{7}, I. M\"uller\altaffilmark{2}, Th. L\"uftinger\altaffilmark{2}, Bruce D. Clarke\altaffilmark{8}, Jennifer R. Hall\altaffilmark{9},  Khadeejah~A.~Ibrahim\altaffilmark{9}}
\altaffiltext{1}{Department of Astronomy, University of Texas, Austin, TX 78712, USA}
\altaffiltext{2}{Institut f\"ur Astronphysik der Universit\"at Wien, T\"urkenschanzstr. 17, A--1180, Wien, Austria}
\altaffiltext{3}{Department of Physical Sciences, Open University, Walton Hall, Milton Keynes MK7 6AA, UK}
\altaffiltext{4}{South African Astronomical Observatory, P.O. Box 9, Observatory 7935, South Africa}
\altaffiltext{5}{Jeremiah Horrocks Institute, University of Central Lancashire, Preston PR1 2HE, UK}
\altaffiltext{6}{Herzberg Institute of Astrophysics, National Research Council of Canada, 5071 West Saanich Road, Victoria, BC V9E 2E7, Canada}
\altaffiltext{7}{N. Copernicus Astronomical Center, Polish Academy of Sciences, ul. Bartycka 18, 00-716 Warszawa, Poland}
\altaffiltext{8}{SETI Institute/NASA Ames Research Center, Moffett Field, CA 94035, USA}
\altaffiltext{9}{Orbital Sciences Corporation/NASA Ames Research Center, Moffett Field, CA 94035, USA}

\begin{abstract} 
Two years of {\it Kepler} data of KIC\,8054146 ($\delta$\,Sct/$\gamma$\,Dor hybrid) revealed 349 statistically significant frequencies between 0.54 and 191.36\,cd$^{-1}$ (6.3\,$\mu$Hz to 2.21\,mHz). The 117 low frequencies cluster in specific frequency bands, but do not show the equidistant period spacings predicted for gravity modes of successive radial order, $n$, and reported for at least one other hybrid pulsator. The four dominant low frequencies in the 2.8 to 3.0\,cd$^{-1}$  (32 to 35\,$\mu$Hz) range show strong amplitude variability with timescales of months and years. These four low frequencies also determine the spacing of the higher frequencies in and beyond the $\delta$\,Sct pressure-mode frequency domain. In fact, most of the higher frequencies belong to one of three families with spacings linked to a specific dominant low frequency. In the Fourier spectrum, these family regularities show up as triplets, high-frequency sequences with absolutely equidistant frequency spacings, side lobes (amplitude modulations) and other regularities in frequency spacings. Furthermore, within two families the amplitude variations between the low and high frequencies are related. We conclude that the low frequencies  (gravity modes, rotation) and observed high frequencies (mostly pressure modes) are physically connected. This
unusual behavior may be related to the very rapid rotation of the star: from a combination of high and low-resolution spectroscopy we determined that KIC\,8054146 is a very fast rotator (\vsini\,=\,300\,$\pm$\,20\,\kms) with an effective temperature
of 7600\,$\pm$\,200\,K and a surface gravity \logg\ of 3.9\,$\pm$\,0.3. Several astrophysical ideas explaining the origin of the relationship between the low and high frequencies are explored.
 \end{abstract} 
 

\section{Introduction}

{\it Kepler} observations have completely changed our understanding of stellar 
pulsations in the $\delta$\,Sct instability strip.  Earlier ground-based
data had shown that $\gamma$~Dor stars, which pulsate in frequencies 
typically below 5\,d$^{-1}$, lie in a fairly small region on, or  just above, 
the main sequence that partly overlaps the cool edge of the $\delta$~Sct 
instability strip.  A few stars were known in which both $\gamma$~Dor and
$\delta$~Sct pulsations were present (the so-called hybrids).  {\it Kepler} photometry 
has shown that practically all A and F stars, whether they are $\delta$~Sct 
stars or not, have low frequencies \citep{Bal11}.  Thus the term `hybrid'
has lost its meaning since nearly all $\delta$~Sct stars are hybrids.  The 
convective blocking mechanism, which is thought to drive $\gamma$~Dor 
pulsations, ceases to be effective for stars hotter than about 7500\,K, yet 
low frequencies are found in the hottest A-type stars.  It is, therefore, not
possible to fully understand the low frequencies in the A-type stars as pulsation.

\begin{figure*}[!htb]
\centering
\includegraphics[bb=30 260 730 520, width=\textwidth, clip]{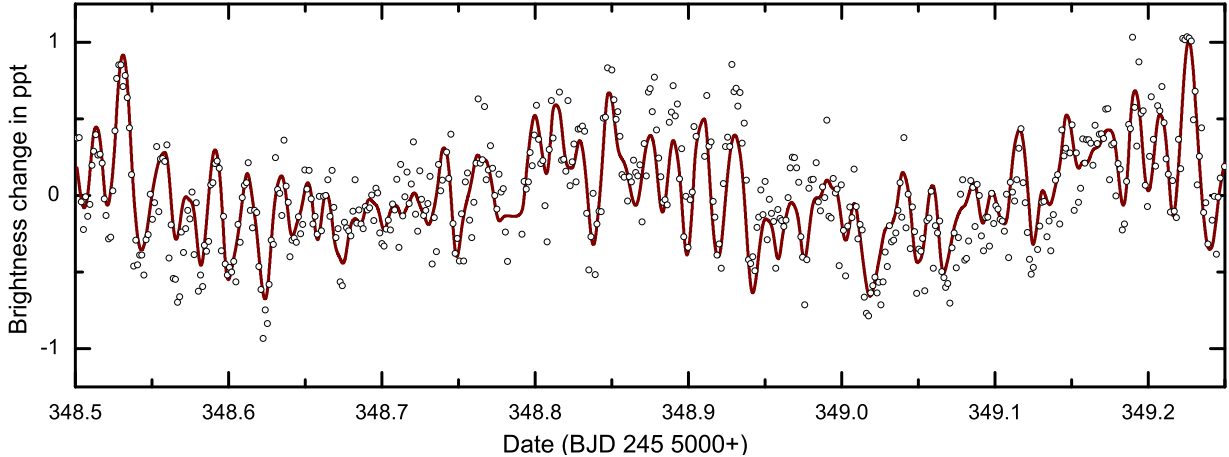}
\caption{Typical light curve covering 0.75 days out of a total of more than 580 days. The plot demonstrates the presence of both low- and high-frequency groups. The average deviation per single point from the fit is 0.18 ppt.}
\label{litecurve}
\end{figure*}

Statistical analyses of the low frequencies in A-type stars show that the
frequencies are compatible with the expected rotational frequencies \citep{Bal11}.
In fact, for the $\delta$\,Sct star KIC\,9700322, a secure identification of low-frequency peaks with stellar rotation was established from a variety of techniques: stellar line width from spectroscopy, the spacing of an observed $\ell$=2 p-mode quintuplet and the time scales of the small amplitude modulation of the dominant modes \citep{BBL11, Breg11}. The low and high frequencies in hybrid pulsators are also related through additional physical mechanisms, which are presently not fully understood. This relationship is seen in a number of $\delta$\,Sct stars measured by space telescopes. An excellent example is the CoRoT star ID 105733033 studied by \citet{C12}, which shows a number of couplings between the low and high frequencies. Another star showing a number of numerical relationships between low and high frequencies is KIC 8054146, the star examined in the present paper. This star is interesting from another point of view as well: its very high rotational velocity (see below).

In general, asteroseismic studies of rapidly rotating stars of spectral type A and early F have been
less successful than those of slow rotators. The reason is both theoretical and observational: rapid rotation
cannot yet be reliably treated in theoretical modeling, while spectroscopically the extreme width of the
spectral lines of the rapidly rotating stars makes detailed abundance analyses very difficult. Furthermore, observational
studies \citep[e.g.,][]{Breg2000} have shown that in $\delta$\,Sct stars low rotation is a prerequisite for high-amplitude radial pulsation, while
the rapidly rotating $\delta$\,Sct stars have small photometric amplitudes with mainly nonradial pulsation.

In this paper we shall study the asteroseismic properties of a very rapidly rotating A/F star, KIC\,8054146, which is also a hybrid pulsator with both $\delta$\,Sct and $\gamma$\,Dor properties \citep{uytterhoeven11}, i.e., a star with reported gravity and pressure modes.
 
\begin{figure*}[!htb]
\centering
\includegraphics[bb=50 30 820 550,width=\textwidth,clip]{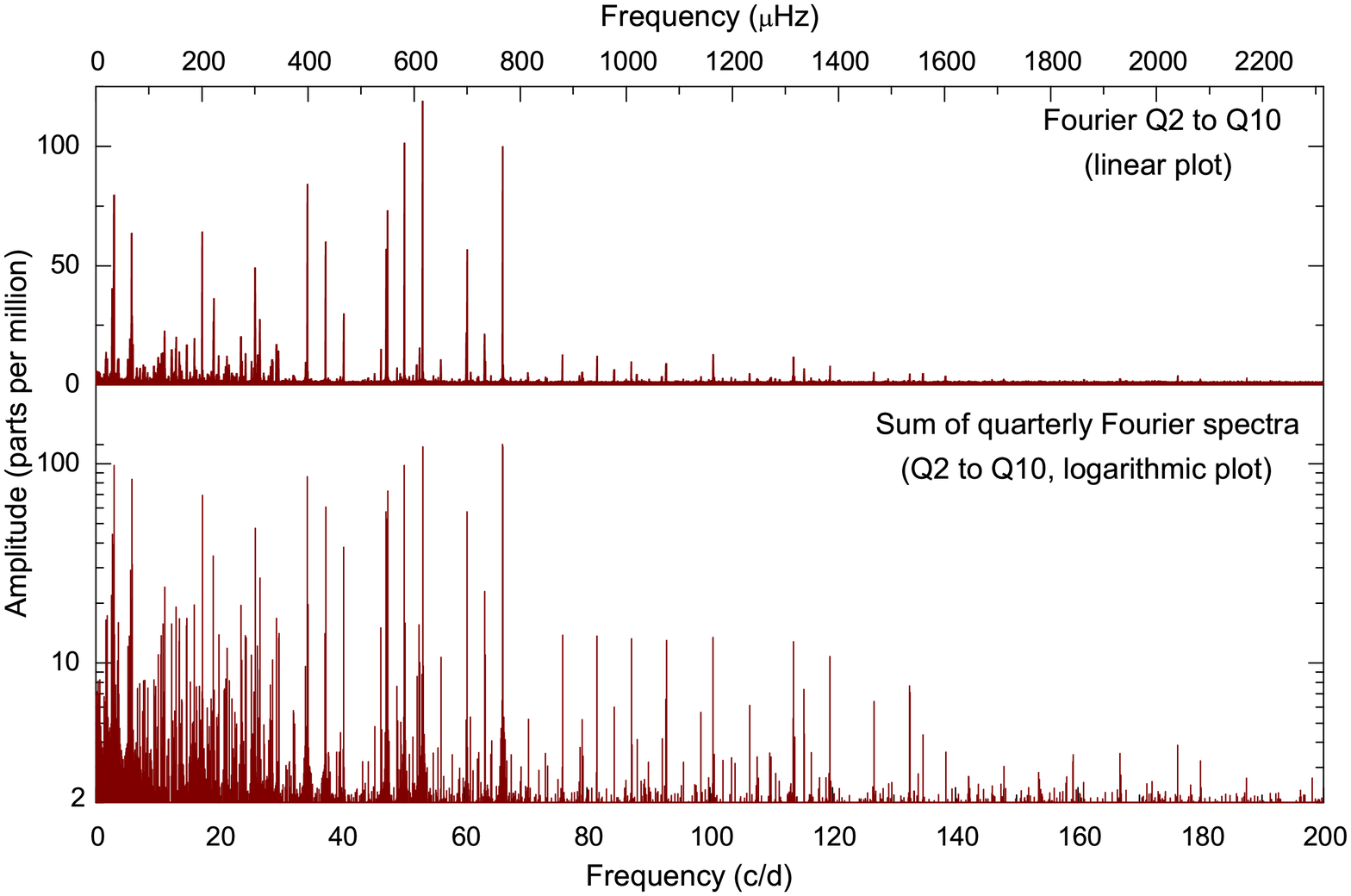}
\caption{The amplitude spectra of KIC\,8054146. The top panel shows the spectrum of the data covering two years. To eliminate the effects of the frequency drifts in the star, the bottom panel uses Fourier spectra covering three months only. Furthermore, in order to demonstrate the many frequencies with small amplitudes, the panel shows amplitudes logarithmically. Note the large frequency range over which the star is variable.}\label{allps}
\end{figure*}

\section{{\it Kepler} observations of KIC\,8054146}
 
The {\it Kepler} mission is designed to detect Earth-like planets around solar-type stars \citep{koch2010}. To achieve that goal, {\it Kepler} is continuously monitoring the brightness of over 150\,000 stars for at least 6\,yr in a 115 square degree fixed field of view. Photometric results show that after one year of almost continuous observations, pulsation amplitudes of 2\,ppm (parts-per-million) are easily detected in the periodogram for stars brighter than $V\,=\,11$\,mag. Two modes of observation are available: long cadence (LC, 29.4-min exposures) and short-cadence (SC, 1-min exposures). With short-cadence exposures it is possible to observe the whole frequency range from 0 to 200\,cd$^{-1}$ seen in $\delta$\,Sct, $\gamma$\,Dor, and hybrid stars pulsating with both the $\delta$\,Sct and $\gamma$\,Dor frequencies. 
 
KIC\,8054146 (K$_p$ = 11.33) was extensively observed with the {\it Kepler} satellite in short cadence during quarters Q2.3 (BJD 245 5064 - 245 5091) as well as Q5 to Q10 (BJD 245 5276 - 245 5833), covering a time span of 27\,d and 557\,d, respectively. The large amount of accurate data covering more than a year leads to an unprecedented frequency resolution, permitting
us to precisely examine the different frequency patterns (separations). The coverage is ideal to examine the relation between the low-frequency variability (possibly caused by gravity modes and stellar rotation) and the higher-frequency $\delta$\,Sct pressure modes.

The {\it Kepler} short-cadence photometry with a measurement per minute is available as calibrated and uncalibrated data. To avoid the known systematic errors present in the calibrated data, we only used the uncalibrated data. We performed our own reductions eliminating the zero-point trends, outliers, jumps and known spurious effects such as those caused by the reaction wheels (spurious frequencies near 0.3\,cd$^{-1}$). Our independently developed reduction process is similar to that published by \citet{garcia2011} for solar-type stars and the description will not be repeated here. Despite the careful reductions, any detections of very low frequencies below 0.3\,cd$^{-1}$ should be treated with caution, but this does not affect the present conclusions. For Q8, a known spurious frequency, drifting from 31.5 to 32.2\,cd$^{-1}$ with amplitudes around 60\,ppm was taken out on a daily basis.

The short-cadence data were supplemented by long-cadence data from quarters Q1 to Q4 (BJD 245 4964 - 245 5275). These measurements of two points per hour were reduced by us in a similar manner to the short-cadence data. A comparison of the resulting LC and SC frequency spectra for the important Q2.3 time period gave an excellent average agreement to $\pm$\,3\,ppm  for the low-frequency peaks, corresponding to the size of the formal uncertainties. Visual comparisons for subsequent quarters with both types of data confirmed that long-cadence data can indeed be used for studies of low-frequency amplitude variability (only). 
This result is in agreement with a discussion of the characteristics of the {\it Kepler} data by \citet{Murphy2012}. Nevertheless, we emphasize that in this paper the LC data are used only to examine the low-frequency amplitude variability from Q1 to Q4.

\section{Frequency analyses}

The {\it Kepler} short-cadence data of KIC\,8054146 were analyzed with the statistical package {\tt PERIOD04}
\citep{LenzBreger2005}. This package carries out multifrequency analyses 
with Fourier as well as least-squares algorithms and the determination of significance of a peak does not rely on the 
assumption of white noise.
 
Following the standard procedures for examining the peaks with {\tt PERIOD04}, we 
have adopted the amplitude signal/noise criterion of 4.0 \citep{B93}. 
This standard technique is modified for all our 
analyses of accurate satellite photometry: the noise is calculated from 
prewhitened data because of the huge range in amplitudes of three orders of 
magnitudes. The extensive data led to very low root-mean-square noise levels in the Fourier spectrum:
1\,ppm (0.5 -- 7\,cd$^{-1}$ range), 0.75\,ppm (7 -- 20\,cd$^{-1}$),  0.6\,ppm (20 -- 30\,cd$^{-1}$),
0.55\,ppm (30 -- 60\,cd$^{-1}$), and 0.5\,ppm (60 -- 200\,cd$^{-1}$).

We undertook extensive computations to eliminate the effects of amplitude and (small) frequency variations, while preserving the extremely high frequency resolution.
A large number of tests involving various amounts of prewhitening of the dominant frequencies were also undertaken with thousands of hours of computations.

We detected 349 statistically significant frequencies between 0.54 and 191.36\,cd$^{-1}$. There is a clear separation in the density of frequency peaks between low and high frequencies: 116 frequencies have values smaller than 6\,cd$^{-1}$, there is a single peak at 6.71\,cd$^{-1}$, while 232 peaks are found between 7 and 192\,cd$^{-1}$ (the high-frequency group). The results are shown in Figure \ref{allps}, which presents the Fourier spectra in two different ways. The top panel ignores the small period and large amplitude variability of all frequencies and presents the average solution. The lower panel represents the average of separate Fourier analyses for each quarter and depicts the result with logarithmic amplitudes in order to emphasize the many low-amplitude peaks.

The frequencies with average amplitudes of 5.0\,ppm or larger are listed in Table \ref{tab_kic}. Due to the strong amplitude variability, these average amplitudes were not computed from a single solution, but were obtained from averaging the amplitudes of the three Q2, Q5-7, and Q8-10 time periods. The Q2 data cover only 27 days, but were given equal weight for determining the averages because of the dominance of f$_1$ and all its related sequences in the early quarters only. On the other hand, the lower frequency resolution of the Q2 data forced us to omit some close frequency pairs from the Q2 solution, explaining a small number of blank spaces in the table. The 120 frequencies in the table can only be a subset of the statistically significant frequencies and by themselves cannot reveal all the rich regularities present in the star. These will be covered below.

Table \ref{tab_kic} also lists a number of identifications such as combination frequencies, harmonics and other sequences, which will be covered below. We have restricted the listed identifications to only the main patterns, mostly involving the dominant frequencies f$_1$ to f$_{16}$. Multiple listings for each frequency component in a combination were also avoided in the interest of clarity. We also note the excellent frequency accuracy:  the predicted and observed frequencies of the combinations fit within a few 0.0001\,cd$^{-1}$.

Are there hundreds or thousands of additional frequencies which we have not detected? If we exclude the low-frequency region, in which the noise is definitely not white, we consider our amplitude limit of 2 to 3 ppm quite restrictive and can rule out a large number of additional modes with photometric amplitudes of this size or larger. In addition, no statistically significant peaks at higher frequencies up to the Nyquist frequency at 368\,cd$^{-1}$ were detected.

\begin{figure}[!htb]
\centering
\includegraphics[bb=30 430 810 950,width=\columnwidth,clip]{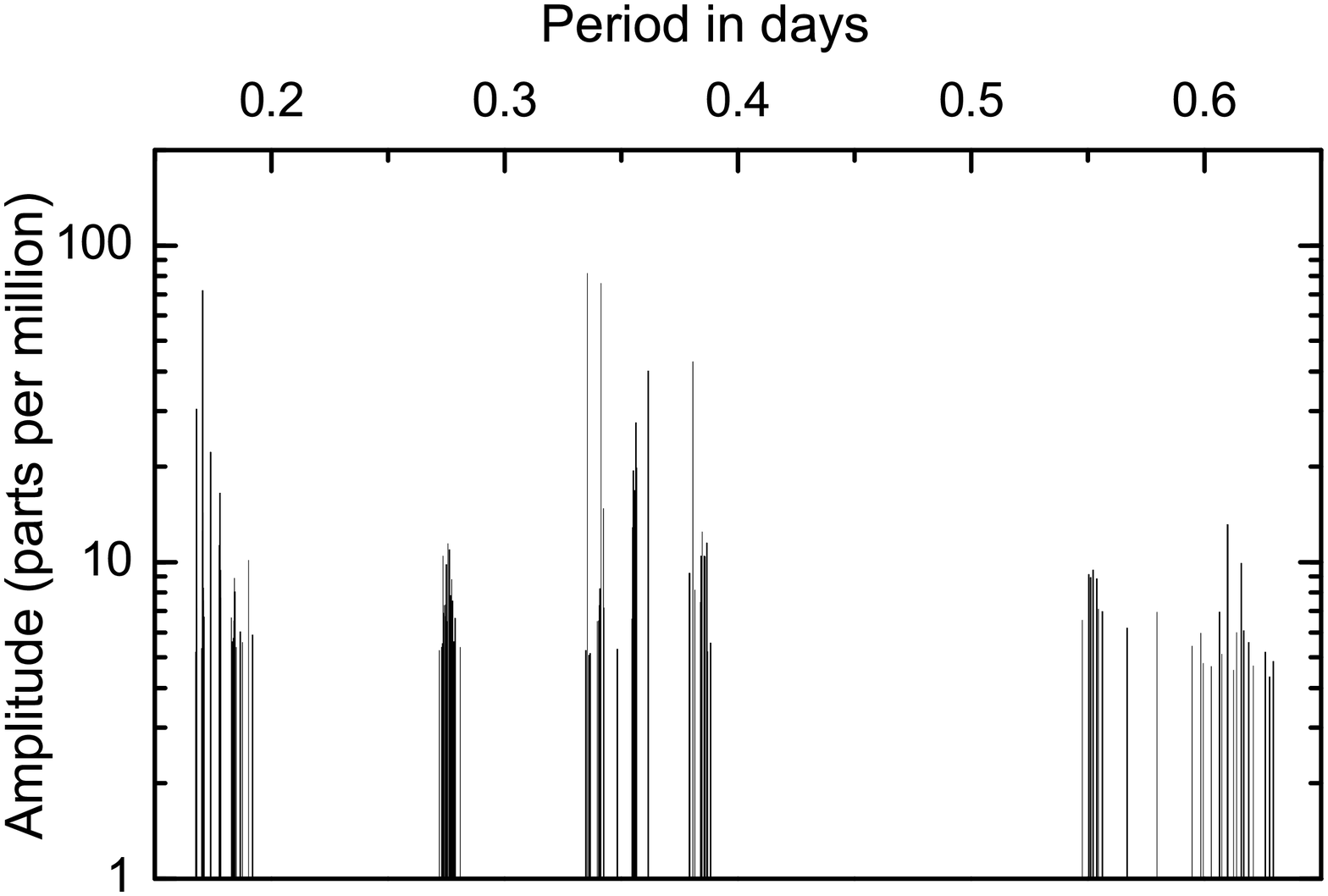}
\caption{The distribution of the low-frequency peaks in period (=\,1/frequency). Period, rather than frequency, is used here since gravity modes are expected to be equally spaced in period, rather than frequency.  Equidistant period spacings have not been detected. Note that the peaks cluster mainly in four specific regions.}
\label{hiper}
\end{figure}

\section{The low frequencies}

The 117 low frequencies detected by us are not randomly distributed. They are mostly clustered in specific frequency bands: 1.6 -- 1.8, 2.6 -- 3.0, 3.5 -- 3.7, and 5.2 -- 6.0\,cd$^{-1}$.
We note that the last two bands are located at approximately twice the frequencies of the first two bands, although only some individual peaks obey a strict f, 2f relation. The
clustering of the peaks is demonstrated in Figure \ref{hiper}. Since gravity modes of adjacent radial order, $n$, are expected to show equal spacing in period in the asymptotic limit, Figure \ref{hiper}
shows the distribution in period (inverse frequency). A cursory examination already reveals that equidistant period (or frequency) spacing is not dominant.

For the hybrid pulsator, CoRoT ID\,105733033, \citet{C12} have discovered a remarkable sequence of asymptotically spaced frequencies with an equidistant period spacing of 0.0307 days.
They interpret this to be caused by $\ell$\,=\,2 gravity modes of different radial order, $n$. Since for KIC\,8054146 we also see a large number of closely spaced peaks at low frequencies, the question arises whether such gravity modes are also present. In KIC\,8054146 the amplitudes of the low-frequency peaks are a factor of 10 -- 100 smaller, but the extensive data set makes such a search possible.

However, a more cautious approach might be in order. Some aliasing exists in our $\it Kepler$ data due to small gaps in coverage. Furthermore, the slow amplitude variability in this star can also lead to close frequencies. These effects need to be minimized for the the low-frequency region with many close frequency peaks.

Consequently, we have repeated our previous analysis with some refinements:
To reduce any effects of aliasing caused by the large gap between Q2 and Q5 as well as the small gaps in the Q5 to Q10 data, we omitted the 27d of the Q2 data.
This reduced the aliasing to four small peaks separated from the main frequencies by less than 0.01\,cd$^{-1}$. To reduce the effects of amplitude variability of the dominant peaks, we
calculated multifrequency solutions for the 10 dominant low frequencies with {\tt PERIOD04} allowing for quarterly amplitude and phase changes.
The prewhitened data were then used to find additional, statistically significant peaks with amplitudes of 5\,ppm or larger.

We then examined the four regions for regular frequency as well as period spacings of the individual peaks by constructing histograms of the observed frequency and period differences. Again, regular patterns could not be detected, in particular, the spacings observed by \citet{C12} in another hybrid star were not found. We conclude that for KIC\,8054146, gravity modes with equidistant period spacings were not detected.

\begin{figure}[!htb]
\centering
\includegraphics[bb=20 12 500 760,width=\columnwidth,clip]{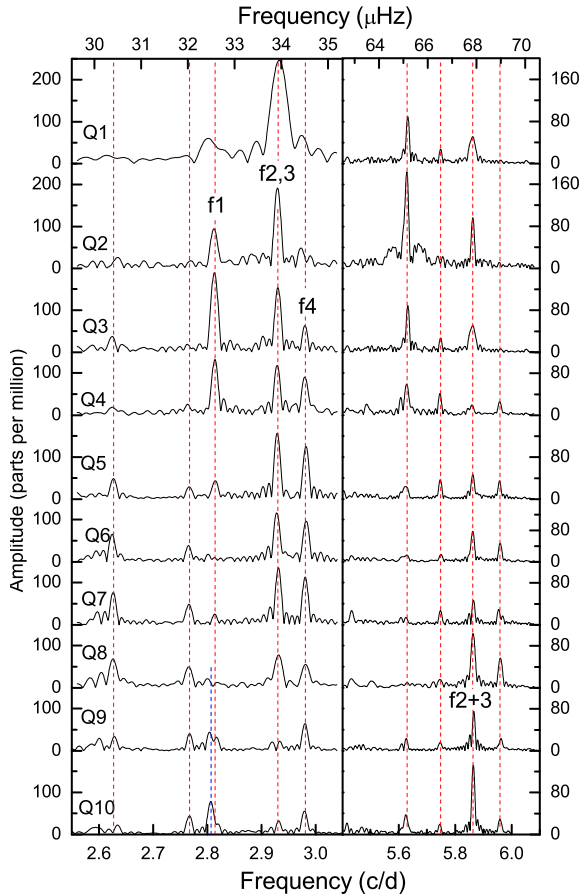}
\caption{Amplitude variability of the dominant low frequencies over 2.5 years. The dominant peaks in 5 -- 6\,cd$^{-1}$ range are mostly multiples and combinations of the 2.5 -- 3.0\,cd$^{-1}$ peaks. 
The amplitude variations of the 2f multiples are not in step with the amplitude variations of f. Note also the strong amplitude decrease of the f$_1$ (2.814\,cd$^{-1}$) peak after Q5 and the ``return'' at 2.806\,cd$^{-1}$ after Q8.}\label{lowamp}
\end{figure}

The four dominant low frequencies in the 2.8 to 3.0\,cd$^{-1}$  (32 to 35\,$\mu$Hz) show strong amplitude variability with time scales of months and years (Figure \ref{lowamp}). An example is the peak at 2.814\,cd$^{-1}$ (f$_1$), which
reaches its maximum value during Q3, slowly decreases its amplitude during Q5 and is almost absent from Q6 to Q9.
The 2f term follows this behavior. The 2.98\,cd$^{-1}$ peak, however, appears in Q3, slowly grows and starts to disappear during Q9 and Q10. The behavior of the 2.93/5.86\,cd$^{-1}$ group of peaks is more complex, and will be discussed below together with its related family of high-frequency modes.

\clearpage
\begin{table}
\begin{scriptsize}
\begin{center}
\onecolumn
\caption{Multifrequency solution of KIC\,8054146, patterns and identifications\label{tab_kic}}
\begin{tabular}{lrrlrrrclrrlrrr}
\tableline\tableline
\noalign{\smallskip}	
\multicolumn{3}{c}{Frequency} & ID& \multicolumn{3}{c}{Amplitudes (ppm)}&\hspace{10mm}& \multicolumn{3}{c}{Frequency} & ID& \multicolumn{3}{c}{Amplitudes (ppm)}\\
& cd$^{-1}$ &$\mu$Hz  & &Q2 & Q5-7 & Q8-10 &\hspace{10mm}& & cd$^{-1}$ &$\mu$Hz & &Q2 & Q5-7 & Q8-10 \\
\noalign{\smallskip}	\hline \noalign{\smallskip}
f$_{1}$	&	2.814	&	32.57	&		&	118	&	10	&	25	&	&	f$_{9}$	&	40.346	&	466.97	&	T$_{1}$, f$_{8}$+f$_{2}$	&	29	&	44	&	35	\\
f$_{2}$	&	2.930	&	33.91	&		&	174	&	122	&	32	&	&	f$_{10}$	&	47.296	&	547.41	&	T$_{3}$, f$_{12}$-f$_{4}$	&	8	&	43	&	90	\\
f$_{3}$	&	2.934	&	33.95	&		&		&	51	&	18	&	&	f$_{11}$	&	47.512	&	549.91	&		&	70	&	74	&	73	\\
f$_{4}$	&	2.981	&	34.50	&		&	35	&	106	&	55	&	&	f$_{12}$	&	50.278	&	581.92	&	T$_{3}$	&	30	&	106	&	112	\\
f$_{5}$	&	17.349	&	200.80	&		&	105	&	89	&	37	&	&	f$_{13}$	&	53.260	&	616.44	&	T$_{3}$, f$_{12}$+f$_{4}$	&	123	&	144	&	98	\\
f$_{6}$	&	25.952	&	300.37	&	T$_{2}$-T$_{1}$	&	18	&	57	&	47	&	&	f$_{14}$	&	60.434	&	699.47	&	T$_{2}$, f$_{15}$-f$_{3}$	&	49	&	43	&	77	\\
f$_{7}$	&	34.482	&	399.10	&	T$_{1}$, f$_{8}$-f$_{3}$	&	31	&	79	&	113	&	&	f$_{15}$	&	63.368	&	733.42	&	T$_{2}$	&	34	&	33	&	8	\\
f$_{8}$	&	37.416	&	433.06	&	T$_{1}$	&	37	&	103	&	25	&	&	f$_{16}$	&	66.297	&	767.33	&	T$_{2}$, f$_{15}$+f$_{2}$	&	188	&	113	&	114	\\
\noalign{\smallskip}
\hline
\noalign{\smallskip}																																							
f$_{17}$	&	1.640	&	18.98	&		&	24	&	9	&	17	&	&	f$_{69}$	&	19.184	&	222.04	&		&	11	&	13	&	66	\\
f$_{18}$	&	2.600	&	30.09	&		&		&	8	&	21	&	&	f$_{70}$	&	19.671	&	227.68	&		&	8	&	2	&	7	\\
f$_{19}$	&	2.627	&	30.40	&		&	27	&	60	&	26	&	&	f$_{71}$	&	20.015	&	231.66	&		&	21	&	12	&	11	\\
f$_{20}$	&	2.766	&	32.01	&	f$_{12}$-f$_{11}$	&	16	&	36	&	42	&	&	f$_{72}$	&	20.048	&	232.04	&		&	10	&	3	&	4	\\
f$_{21}$	&	3.620	&	41.89	&		&	2	&	16	&	7	&	&	f$_{73}$	&	20.856	&	241.39	&		&	8	&	1	&	9	\\
f$_{22}$	&	5.629	&	65.15	&	2f$_{1}$	&	79	&	9	&	14	&	&	f$_{74}$	&	20.980	&	242.82	&S$_{30}$-f$_{16}$		&	21	&	5	&	3	\\
f$_{23}$	&	5.745	&	66.49	&	f$_{1}$+f$_{2}$	&	18	&	4	&	5	&	&	f$_{75}$	&	21.096	&	244.16	&	S$_{29}$-f$_{15}$	&	5	&	11	&	3	\\
f$_{24}$	&	5.864	&	67.87	&	f$_{2}$+f$_{3}$	&	111	&	50	&	88	&	&	f$_{76}$	&	21.156	&	244.86	&		&	11	&	8	&	8	\\
f$_{25}$	&	5.867	&	67.91	&	2f$_{3}$	&		&	8	&	27	&	&	f$_{77}$	&	21.324	&	246.80	&		&	9	&	7	&	2	\\
f$_{26}$	&	5.958	&	68.95	&		&	6	&	35	&	25	&	&	f$_{78}$	&	21.360	&	247.22	&		&	5	&	16	&	10	\\
f$_{27}$	&	7.174	&	83.03	&	f$_{14}$-f$_{13}$	&	10	&	6	&	10	&	&	f$_{79}$	&	21.744	&	251.66	&		&	4	&	10	&	8	\\
f$_{28}$	&	7.616	&	88.15	&		&	9	&	8	&	2	&	&	f$_{80}$	&	22.165	&	256.54	&		&	4	&	2	&	9	\\
f$_{29}$	&	7.740	&	89.59	&		&	4	&	7	&	11	&	&	f$_{81}$	&	22.690	&	262.62	&		&	15	&	2	&	3	\\
f$_{30}$	&	7.986	&	92.43	&		&	13	&	10	&	5	&	&	f$_{82}$	&	23.657	&	273.81	&		&	7	&	24	&	18	\\
f$_{31}$	&	8.409	&	97.32	&		&	20	&	8	&	2	&	&	f$_{83}$	&	23.788	&	275.32	&	S$_{31}$-f$_{16}$	&	14	&	11	&	9	\\
f$_{32}$	&	8.443	&	97.72	&	3f$_1$	&	7	&	5	&	3	&	&	f$_{84}$	&	24.398	&	282.39	&		&	16	&	16	&	10	\\
f$_{33}$	&	9.497	&	109.92	&		&	10	&	7	&	8	&	&	f$_{85}$	&	25.331	&	293.19	&		&	18	&	8	&	10	\\
f$_{34}$	&	9.724	&	112.55	&	S$_{26}$-f$_{16}$	&	17	&	5	&	4	&	&	f$_{86}$	&	25.709	&	297.55	&		&	6	&	9	&	7	\\
f$_{35}$	&	10.108	&	116.99	&	f$_{15}$-f$_{13}$	&	12	&	12	&	2	&	&	f$_{87}$	&	25.721	&	297.70	&		&	8	&	4	&	8	\\
f$_{36}$	&	10.156	&	117.54	&	f$_{14}$-f$_{12}$	&	3	&	9	&	15	&	&	f$_{88}$	&	25.935	&	300.17	&		&	7	&	2	&	11	\\
f$_{37}$	&	10.546	&	122.06	&		&	17	&	8	&	9	&	&	f$_{89}$	&	26.375	&	305.26	&		&	6	&	14	&	12	\\
f$_{38}$	&	10.674	&	123.54	&		&	15	&	19	&	6	&	&	f$_{90}$	&	26.610	&	307.99	&	S$_{32}$-f$_{16}$	&	33	&	6	&	5	\\
f$_{39}$	&	10.809	&	125.11	& 	&	9	&	3	&	4	&	&	f$_{91}$	&	26.743	&	309.52	&		&	14	&	31	&	25	\\
f$_{40}$	&	10.827	&	125.31	&		&	9	&	3	&	6	&	&	f$_{92}$	&	28.457	&	329.36	&		&	6	&	9	&	7	\\
f$_{41}$	&	10.916	&	126.34	&		&	24	&	14	&	12	&	&	f$_{93}$	&	29.431	&	340.63	&	S$_{33}$-f$_{16}$	&	14	&	20	&	14	\\
f$_{42}$	&	11.042	&	127.80	&		&	9	&	7	&	1	&	&	f$_{94}$	&	29.801	&	344.92	&		&	6	&	16	&	13	\\
f$_{43}$	&	11.211	&	129.76	&		&	26	&	30	&	14	&	&	f$_{95}$	&	32.239	&	373.14	&	S$_{34}$-f$_{16}$	&	17	&	6	&	1	\\
f$_{44}$	&	11.260	&	130.32	&4f$_1$:		&	25	&	3	&	9	&	&	f$_{96}$	&	32.355	&	374.48	&	S$_{33}$-f$_{15}$	&	14	&	4	&	2	\\
f$_{45}$	&	11.763	&	136.15	&		&	8	&	3	&	4	&	&	f$_{97}$	&	34.185	&	395.66	&	S$_{29}$-f$_{12}$	&	2	&	9	&	11	\\
f$_{46}$	&	12.366	&	143.12	&		&	23	&	17	&	12	&	&	f$_{98}$	&	45.444	&	525.97	&	S$_{33}$-f$_{12}$	&		&	3	&	7	\\
f$_{47}$	&	12.860	&	148.85	&		&	8	&	4	&	6	&	&	f$_{99}$	&	46.488	&	538.05	&		&	17	&	16	&	14	\\
f$_{48}$	&	13.037	&	150.89	&	f$_{16}$-f$_{13}$	&	31	&	19	&	15	&	&	f$_{100}$	&	49.094	&	568.22	&		&	4	&	3	&	12	\\
f$_{49}$	&	13.090	&	151.50	&	f$_{15}$-f$_{12}$	&	3	&	32	&	10	&	&	f$_{101}$	&	49.608	&	574.17	&		&	7	&	5	&	5	\\
f$_{50}$	&	13.604	&	157.46	&		&	25	&	14	&	12	&	&	f$_{102}$	&	52.326	&	605.62	&		&	10	&	8	&	9	\\
f$_{51}$	&	13.655	&	158.04	&		&	8	&	10	&	6	&	&	f$_{103}$	&	52.694	&	609.88	&		&	13	&	15	&	15	\\
f$_{52}$	&	13.662	&	158.12	&		&		&	10	&	5	&	&	f$_{104}$	&	56.236	&	650.88	&	f$_{12}$+f$_{26}$	&	7	&	13	&	8	\\
f$_{53}$	&	13.972	&	161.72	&		&	7	&	5	&	6	&	&	f$_{105}$	&	61.088	&	707.04	&		&	5	&	5	&	5	\\
f$_{54}$	&	14.015	&	162.21	&		&	7	&	9	&	4	&	&	f$_{106}$	&	63.427	&	734.10	&		&	14	&	10	&	9	\\
f$_{55}$	&	14.045	&	162.56	&		&	18	&	0	&	5	&	&	f$_{107}$	&	76.022	&	879.89	& S$_{26}$		&	23	&	15	&	9	\\
f$_{56}$	&	14.802	&	171.32	&		&	16	&	14	&	20	&	&	f$_{108}$	&	81.649	&	945.01	& S$_{28}$		&	27	&	11	&	11	\\
f$_{57}$	&	15.351	&	177.67	&	S$_{28}$-f$_{16}$	&	22	&	6	&	2	&	&	f$_{109}$	&	87.279	&	1010.17	&	S$_{30}$	&	26	&	15	&	7	\\
f$_{58}$	&	15.855	&	183.51	&	f$_{15}$-f$_{11}$	&	15	&	5	&	1	&	&	f$_{110}$	&	92.908	&	1075.32	&	S$_{32}$	&	33	&	12	&	6	\\
f$_{59}$	&	16.019	&	185.41	&	f$_{16}$-f$_{12}$	&	7	&	20	&	22	&	&	f$_{111}$	&	98.536	&	1140.46	&	S$_{34}$	&	15	&	6	&	2	\\
f$_{60}$	&	16.078	&	186.09	&		&	24	&	10	&	5	&	&	f$_{112}$	&	100.557	&	1163.86	& 2f$_{12}$	&	3	&	12	&	14	\\
f$_{61}$	&	16.392	&	189.72	&		&	19	&	7	&	4	&	&	f$_{113}$	&	106.520	&	1232.87	&	2f$_{13}$	&	6	&	6	&	3	\\
f$_{62}$	&	16.865	&	195.19	&		&	20	&	0	&	6	&	&	f$_{114}$	&	113.646	&	1315.35	&	f$_{12}$+f$_{15}$	&	10	&	18	&	6	\\
f$_{63}$	&	17.080	&	197.68	&		&	17	&	1	&	5	&	&	f$_{115}$	&	113.694	&	1315.90	&	f$_{13}$+f$_{14}$	&	8	&	4	&	5	\\
f$_{64}$	&	18.165	&	210.24	&	S$_{29}$-f$_{16}$	&	5	&	4	&	7	&	&	f$_{116}$	&	115.393	&	1335.57	&		&	6	&	2	&	12	\\
f$_{65}$	&	18.785	&	217.42	&	f$_{16}$-f$_{11}$	&	7	&	6	&	7	&	&	f$_{117}$	&	119.558	&	1383.77	&	f$_{13}$+f$_{16}$	&	16	&	11	&	9	\\
f$_{66}$	&	19.001	&	219.92	&	f$_{16}$-f$_{10}$	&	4	&	4	&	7	&	&	f$_{118}$	&	126.731	&	1466.80	&	f$_{14}$+f$_{16}$	&	4	&	4	&	8	\\
f$_{67}$	&	19.106	&	221.14	&		&	9	&	4	&	3	&	&	f$_{119}$	&	132.595	&	1534.66	&	2f$_{16}$	&	18	&	4	&	7	\\
f$_{68}$	&	19.172	&	221.90	&		&		&	1	&	16	&	&	f$_{120}$	&	159.206	&	1842.66	&	S$_{32}$+f$_{16}$	&	12	&	2	&	1	\\
\noalign{\smallskip}	\tableline
\end{tabular}
\end{center}
T: triplets, S: equidistant picket-fence series. The table contains only frequencies with amplitudes of $\geq$ 5.0\,ppm.\\
Almost all amplitudes are variable and the averages are listed. Formal errors are 2\,ppm or smaller.
\end{scriptsize}
\end{table}
\clearpage

\twocolumn

\section{The high frequencies}

We have detected 232 statistically significant frequency peaks between 7 and 191\,cd$^{-1}$. This large range in frequency is unusual for $\delta$\,Sct stars. Not surprisingly, some of the frequency peaks are expected linear combinations of high frequencies and harmonic terms, but cannot explain most of the high frequencies.

The dominant high frequencies are three groups of triplets (T$_1$ to T$_3$), the difference between the T$_1$ and T$_2$ triplets (=\,25.952\,cd$^{-1}$), and a `lonely' peak at 17.349\,cd$^{-1}$.
The triplets show very strong amplitude variability with a timescale of a year or longer. The most striking property of the detected frequency peaks is their separation in frequency,
which often is identical to the previously detected low-frequency peaks.

\begin{figure}[!htb]
\centering
\includegraphics[bb=50 0 810 550,width=\columnwidth,clip]{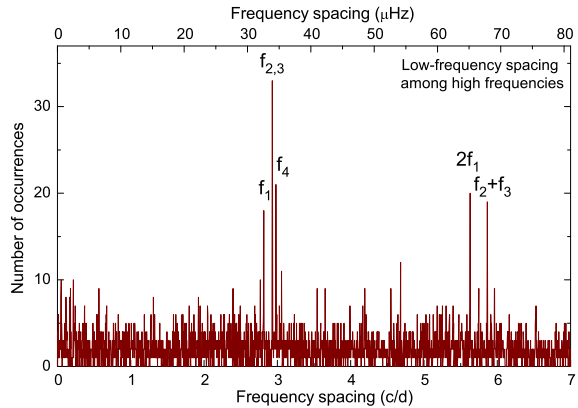}
\caption{Histogram of the frequency spacings among the high frequencies (7 to 200\,cd$^{-1}$) using a narrow bin size of 0.0033\,cd$^{-1}$. The most common spacings are found to be numerically equal to the dominant low frequencies, which have been marked in the diagram. The diagram shows that the low frequencies in the g-mode region strongly affect the spacings of a large number of high frequencies in the p-mode domain.}\label{fcyspacing}
\end{figure}

A powerful method to look for regularities in frequency spacing consists of forming all possible frequency differences between all significant frequency peaks and to examine the histograms. The huge number of
computed differences has to lead to some accidental agreements. In the present case, the high accuracy of the detected frequencies permits us to choose extremely narrow histogram boxes.
Furthermore, we restrict the analysis to spacings smaller than 7\,cd$^{-1}$. We have calculated all the frequency spacings between the 232 high-frequency peaks between 7 and 200\,cd$^{-1}$.
A histogram was computed with a very small box size of 0.0033\,cd$^{-1}$.  The resulting histogram with 2100 boxes is shown in Figure \ref{fcyspacing}. We regard the result as remarkable.

The four dominant low frequencies, f$_1$ to f$_4$, account for 136 frequency spacings among the 232 high-frequency peaks. A few of these spacings (within the narrow adopted range of only 0.0033\,cd$^{-1}$) may be accidental. However, we believe the real number of exact spacings corresponding to observed low frequencies to be even higher because of our restriction to only simple agreements,
e.g., spacings involving the (f$_1$+f$_4$) sum are ignored by our technique.

The preferred spacings involving f$_1$ to f$_4$ can be easily understood in terms of three families of pulsation modes. Each family behaves differently and may require different astrophysical explanations.

\subsection{The f$_2$, f$_3$ family and triplets}

Figure \ref{triplets} shows the two triplets, T$_1$ and T$_2$, and their relationship to three low frequencies and the 25.9517\,cd$^{-1}$ peak. All shown peaks have high amplitudes with a corresponding high accuracy of the frequency
values ($\pm$\,0.0002\,cd$^{-1}$). While data covering only one quarter would suggest equal separations, the two-year data made it possible to detect f$_2$ and f$_3$, which are clearly separated by 0.0040\,$\pm$\,0.0003\,cd$^{-1}$. This small difference has strong implications. We note:

\begin{figure}[!htb]
\centering
\includegraphics[bb=70 80 760 615,width=\columnwidth,clip]{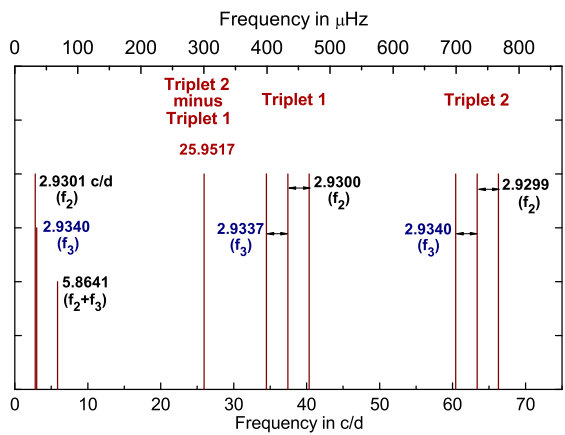}
\caption{The spacings of the two high-frequency triplets and their relation to dominant low frequencies. The two triplets are separated by 25.9517\,cd$^{-1}$, which is also detected as a separate frequency peak. Also, the spacings between the triplets correspond to a close low-frequency pair and the sum of that pair. The separate T$_1$ and T$_2$ triplet components share similar amplitude changes.}\label{triplets}
\end{figure}

(i) The two triplets have nearly identical properties, including amplitude variability. Their frequency separation of 25.9517\,cd$^{-1}$ is also seen as a strong separate peak.

(ii) The frequency differences between the central and right peaks correspond to an observed, dominant low frequency peak at 2.9300\,cd$^{-1}$ (f$_2$).

(iii) The frequency differences between the central and left peaks correspond to an observed, dominant low frequency peak at 2.9340\,cd$^{-1}$ (f$_3$).

(iv) The frequency differences between the left and right peaks correspond to an observed, dominant low frequency peak at 5.8641\,cd$^{-1}$ (f$_2$+f$_3$). This peak is $not$ a harmonic of either
f$_2$ or f$_3$.

(v) The amplitude changes of the f$_2$, f$_3$, and (f$_2$+f$_3$) peaks mirror the amplitude changes of the individual components of the triplets. The $decrease$ of the f$_2$ and f$_3$ strengths after Q7 are
closely mirrored by the decrease of the amplitudes of the central triplets components. On the other hand, the $increase$ of the (f$_1$+f$_2$) amplitude after Q7 is mirrored by the increase in the non-central components
of the triplets. (The lengthy detailed analysis of this phenomenon, which includes analyses of weekly amplitude and phase changes, will be the subject of a later report.)

The observed measurements are incompatible with an interpretation that the 5.86\,cd$^{-1}$ peak is a 2f harmonic of a mode near 2.93\,cd$^{-1}$ and may challenge such interpretations in other hybrid stars.
In Section \ref{discussion} we will explore the possibility that the triplets are the counterparts of the $\ell$\,=\,1 triplets commonly seen in slowly rotating stars and the possibility of combination modes.

\subsection{Equidistant S sequences involving f$_1$}

Between 76 and 102\,cd$^{-1}$, the bottom panel of Figure \ref{allps} shows five equidistantly spaced peaks with additional smaller peaks exactly halfway between them. The separation is exactly equal to the dominant low
frequency, f$_1$, at 2.814\,cd$^{-1}$. The frequencies of the sequence are not exact multiples of f$_1$, but obey a relation of f\,=\,f$_0$\,+\,$n$ f$_1$, where f$_0$\,=\,2.852, f$_1$\,=\,2.8142, and $n$ ranges from
25 to 34. (Note that the high precision of the spacing, f$_1$, is achieved by dividing the detected high frequencies by $n$. Furthermore, f$_0$ is the mathematical intercept of the relation and not a detected peak.)
We shall adopt the relation for our notation, e.g., the frequency at 92.908\,cd$^{-1}$ has $n$\,=\,32 and is called S$_{32}$.

There are several identical S sequences: the sequences are also present as side lobes of the dominant modes. As an example, the dominant mode at f$_{16}$\,=\,66.297\,cd$^{-1}$ shows side lobes
at 26.610 (actually -26.610) and 159.206\,cd$^{-1}$, corresponding to f$_{16}$-S$_{32}$ and f$_{16}$+S$_{32}$, respectively. There are a large number of additional agreements for all the dominant modes and S$_{25}$ to S$_{35}$.
We emphasize that the large number of exact agreements between the observed and computed frequencies rules out accidental agreements and subsequent overinterpretation.

The S sequences are particularly strong during Q2, in which f$_1$ also has a high amplitude. In subsequent quarters, the amplitudes of both f$_1$ and the S sequences decrease. This confirms the
astrophysical connection between the low and high frequencies.

\subsection{Triplet 3 and combinations with f$_4$}

There exists a third connection between a dominant low frequency and the spacings of high frequencies. The triplet T$_3$ between 47 and 53\,cd$^{-1}$
has frequency separations of exactly f$_4$ (2.981\,cd$^{-1}$).
Furthermore, various low-amplitude modes also show frequency differences corresponding to exactly f$_4$.

\section{Spectroscopic observations}

Three spectra of KIC\,8054146 were obtained on 2011 June 29, July 16 and
September 6 with the High Resolution Spectrograph (HRS) attached to the 9.2-m
Hobby-Eberly Telescope (HET) at the McDonald observatory. This is a
cross-dispersed \'echelle spectrograph yielding a resolving power ($R$) 
of 60\,000. The signal-to-noise ratio (S/N) per pixel at 
$\lambda\sim$5000\,\AA\ is 70, 82, and 75, respectively.

The basic data reduction was carried out using the IRAF\footnote{IRAF is
 distributed by the National Optical Astronomy Observatories, which
 are operated by the Association of Universities for Research in
 Astronomy, Inc., under cooperative agreement with the National
 Science Foundation.} software suite.  Bias subtraction was performed 
by removing an averaged zero frame for each night, and spectral images
of a flat-field lamp taken with the same setup as used for our target
observations were used to correct uneven pixel response across the two 
CCDs.  Wavelength calibration (accurate to $\sim 6 \times 10^{-5}$\,\AA) was derived 
from a ThAr emission lamp observed shortly before or after our stellar
spectra were taken.  Cosmic rays were identified as outliers to 2D fits 
to the spectral orders and removed.  In rare cases where a cosmic ray
fell across the \'echelle order, defeating our normal rejection
routine, we have removed it with IRAF's COSMICRAYS utility.  The
finalized spectrum for each night was produced by coadding the three cosmic 
ray splits, resulting in a total effective exposure time of 60 minutes.

After the application of the heliocentric correction, we checked that no
radial velocity variation was present among the three acquired spectra. This excludes
the possibility that the star is a member of a close binary system. This allowed 
us to average the three acquired spectra to increase the S/N, reaching a S/N 
of 131 for the final spectrum. 

The average spectrum, normalized by fitting a low order polynomial to 
carefully selected continuum points, covers the wavelength range 
4100--7900\,\AA, with a gap between 5927\,\AA\ and 6024\,\AA, because one 
\'echelle order is lost in the gap between the two chips of the CCD mosaic 
detector. Given the very high projected rotational velocity (\vsini) of the 
star, we adopted the hydrogen lines as primary indicators for the effective 
temperature \Teff\ determination. The normalization of hydrogen lines, 
observed with \'echelle spectrographs, gives the largest contribution to the 
total error budget in the \Teff\ determination by hydrogen line fitting. To 
reduce this uncertainty we normalized the H$\alpha$ and H$\beta$ lines 
individually for each of the three spectra, obtaining in this way six 
quasi-independent measurements of \Teff. We normalized the hydrogen lines 
making use of the artificial flat-fielding technique described in 
\citet{barklem02}. 

To compute model atmospheres, we employed the \llm\ stellar model atmosphere 
code \citep{llm}. For all the calculations, local thermodynamical equilibrium 
(LTE) and plane-parallel geometry were assumed. We used the VALD database 
\citep{vald1,vald2,vald3} as a source of atomic line parameters for 
opacity calculations with the \llm\ code. Finally, convection was implemented 
according to the \citet{cm1,cm2} model of convection. For more details see \citet{heiter}.

We performed the \Teff\ determination by fitting synthetic line profiles, 
calculated with \synth\ \citep{synth3}, to the observed H$\alpha$ and H$\beta$
line profiles. For each spectrum we derived that the best fitting \Teff\ is 
7600\,K for the H$\alpha$ line and 7700\,K for the H$\beta$ line, with a 
typical uncertainty of 150\,K.

To decrease even more the uncertainty on the effective temperature due to the
normalization, on 2011 September 2 we obtained a spectrum of KIC\,8054146 
centered on the H$\alpha$ line with the Cassegrain low-resolution spectrograph 
attached to the 1.8-m telescope of the Dominion Astrophysical Observatory 
(DAO), Canada. The adopted configuration of the spectrograph provided a 
resolving power of about 15\,000 and the obtained spectrum, covering the 
6480 -- 6780\,\AA\ wavelength region, has a S/N per pixel of 60, calculated 
at about 6600\,\AA. Since the spectrum covers the entire H$\alpha$ line 
profile within a single order, the normalization line is a simple first 
order polynomial which can be safely determined by using the available 
continuum regions on either side of the hydrogen line. This decreases greatly 
the uncertainty on the measurement of \Teff\ due to the continuum 
normalization. By fitting synthetic line profiles to this low-resolution 
spectrum, we derived a \Teff\ of 7500\,$\pm$\,200\,K, where the uncertainty is 
mostly due to the quality of the spectrum.

\begin{figure}[!htb]
\begin{center}
\includegraphics[width=\columnwidth,clip]{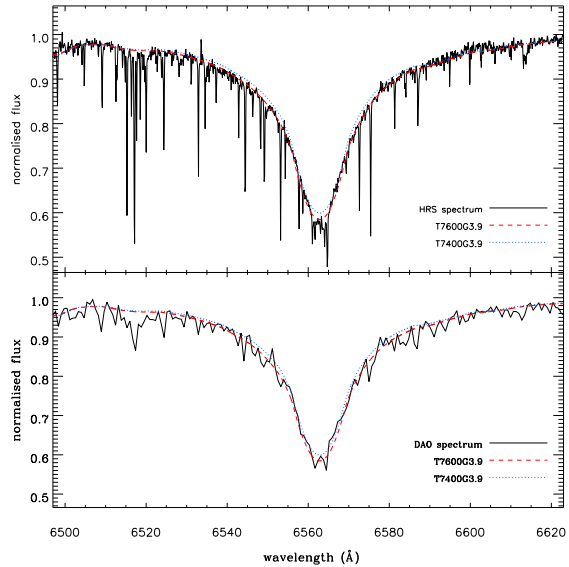}
\caption{Top panel: comparison between the H$\alpha$ line profile (black solid
line), observed with HRS, and synthetic profiles calculated with our best
fitting effective temperature of 7600\,K (red dashed line) and with an 
effective temperature of 7400\,K (blue dotted line). For both synthetic spectra
we adopted \logg \,=\,3.9, \vmic \,=\,2.5\,\kms, and solar metallicity. Bottom panel: as
in the top panel, but with the spectrum observed at the DAO observatory.
The sharp lines visible in the observed spectra are telluric spectral lines;
the stellar photospheric spectral lines are much shallower due to the high rotation.} 
\label{fig:hydrogen} 
\end{center} 
\end{figure}

By averaging all these \Teff\ measurements, we finally 
adopted \Teff\,=\,7600\,K with a formal uncertainty of 80\,K, which is
not realistic, as described later. Figure~\ref{fig:hydrogen} shows a comparison
between the observed HRS and DAO spectra with synthetic spectra, calculated 
with the adopted \Teff\ of 7600\,K and a \Teff\ of 7400\,K, for comparison.
In the procedure of fitting synthetic hydrogen line profiles to the observed 
spectra, we adopted a fixed value of \logg\,=\,3.9 (cgs) for the surface gravity and 
of 2.5\,\kms\ for the microturbulence velocity (\vmic). All model atmospheres 
were produced assuming the solar chemical composition \citep{asplund09}. With 
this set of fundamental parameters, by fitting the most prominent blends we produced synthetic spectra of the 
available spectral region and measured a radial velocity (\vr) of 
31\,$\pm$\,5\,\kms\ and a \vsini\ of 300\,$\pm$\,20\,\kms.

Following the procedure described in \citet{fossati07}, we attempted a 
measurement of the iron abundance from a few \ion{Fe}{1} and \ion{Fe}{2} 
lines, but the limited S/N of the spectrum, the extreme blending and the 
uncertainty on the continuum normalization prevented us from obtaining any 
useful result. As a consequence, we were not able to determine any reliable 
\logg\ value by means of the ionisation equilibrium. At the temperature of 
KIC\,8054146, the two analysed hydrogen lines display a little reaction to 
gravity variations, which allowed us to determine at least an upper and lower 
\logg\ value of 4.2 and 3.4, respectively. The best fitting \logg\ value is 
3.9, indicating that the star is likely to be still on the main-sequence. A
more precise \logg\ value could be obtained either with narrow band
photometry\footnote{\citet{uytterhoeven11} published values of \Teff\ and 
\logg\ obtained from Str\"omgren photometry, but a number of available narrowband measurements of this star have
been found to be those of a nearby brighter star (TYC 3149-1743-1).}, or with a distance 
value, or with calibrated photometry spread over the region of the Balmer 
jump, but none of this is currently available. A more realistic value of 
\errTeff\ can then be obtained by taking into account the limits given 
for \logg, leading to \errTeff\,=\,200\,K. 

\begin{figure}[!htb]
\begin{center}
\includegraphics[width=\columnwidth,clip]{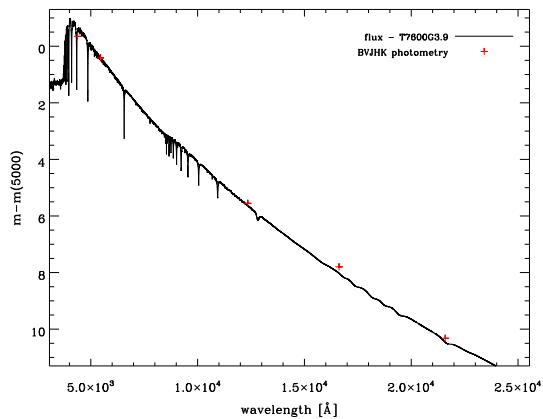}
\caption{Comparison between \llm\ theoretical fluxes calculated with the 
fundamental parameters derived for KIC\,8054146 with Johnson $BV$ and 2MASS
$JHK$ photometry.} 
\label{fig:sed} 
\end{center} 
\end{figure}

The large uncertainties on the abundance values derived from the most
prominent blends made it also impossible to determine a reliable \vmic\ value,
which we therefore kept fixed at 2.5\,\kms, typical of late A-type stars 
\citep[see e.g.][]{fossati08}. Figure~\ref{fig:sed} shows a comparison between the \llm\ theoretical fluxes,
calculated with \Teff\,=\,7600\,K and \logg\,=\,3.9, and Johnson $BV$ and 2MASS $JHK$
photometry. This comparison confirms that the derived \Teff\ fits well the 
available broad-band photometry and the absence of infrared excess excludes the
presence of warm dust around the star. 

In conclusion, the spectroscopic analysis allowed us to determine that 
KIC\,8054146 is a very fast rotating late A-type star, which is likely to have 
a metallicity close to that of the Sun.

\section{Discussion}\label{discussion}

We have shown from the {\it Kepler} data that the low frequencies (up to 7\,cd$^{-1}$) and the high frequencies (up to 200\,cd$^{-1}$) are related in several families through preferred frequency spacings, frequency patterns, and similar amplitude variations in the individual families. Consequently, an explanation in terms of independent gravity and pressure modes appears too simple. The analyses of stars of spectral type A and F with the {\it Kepler} spacecraft have shown that the light output is modulated
by the rotational frequency \citep{Bal11}.
Even a very `simple' A star with radial pulsation modes and a very low rotation rate of $\sim$ 20\,\kms\ already shows complex rotational variations including small amplitude modulations of the radial modes \citep{BBL11}.
Consequently, a very rapidly rotating star such as KIC\,8054146, presumably with differential rotation, might be very interesting. What is the rotation frequency of this star?

Based on the mean values of \Teff\ (7600\,K) and \logg\ (3.9) measured in the previous section we obtain a radius of about 1.9\,R$_\odot$ for
the star (assuming a mass of 1.8\,M$_{\odot}$). Together with the mean measured value of the projected rotational velocity, \vsini\,=\,300 \kms\ and assuming equator-on view ($i$\,=\,90$^{\rm o}$)
we derive a rotational frequency of 2.87\,cd$^{-1}$ (corresponding to 0.75 of the Keplerian breakup rate). Taking into account the given uncertainty of 20\,\kms\ in \vsini,
we obtain a rotational frequency of 2.73\,cd$^{-1}$ (i.e., 0.69 of Keplerian breakup rate) for the lower limit of \vsini\,=\,280\,\kms.
The rotational frequency is, therefore, very similar to that of the dominant low frequencies between 2.8 and 3.0\,cd$^{-1}$,
but the uncertainties exclude the possibility of identifying any one of the peaks as the rotational frequency.

Due to the high value of the surface rotation velocity near breakup velocity, we can conclude that it is more likely that the star is viewed nearly equator-on.
This fact helps us to reduce the uncertainties in the fundamental parameters, since, due to gravity darkening, parameters such as the effective temperature
are a function of aspect in rapidly rotating stars. 
Consequently, the measured temperature and gravity values
are essentially the equatorial values. At the poles, we expect gravity brightening, since both the \Teff\ and \logg\ values are higher there.

\subsection{Rotationally split $\ell$\,=\,1 triplets?}

We do not have a physical explanation for this perfect link between the low frequencies and the higher-frequency triplets, but can offer a speculation:
the triplets are the counterparts of the rotationally split $\ell$\,=\,1 triplets commonly observed in slowly rotating stars. KIC\,8054146, however, is an extremely fast rotator
for which the amount of asymmetry in the frequency separations is not known. An effect that could make the frequency separation between rotationally split modes symmetric
is the 1:1:1 resonance \citep{Buch97}. Such explanations in terms of resonances must, however, be regarded as speculative at this time.

It is important to point out that the small, but statistically significant, departure from equidistance between the components of the two triplets, T$_1$ and T$_2$,
is almost impossible to discover in stars not measured as extensively as KIC\,8054146.
The separations are 2.9300\,$\pm$\,0.0001\,cd$^{-1}$ and 2.9339\,$\pm$\,0.0002\,cd$^{-1}$, respectively.
Even more remarkable is the fact that both these separations are seen as distinct low-frequency peaks and as a sum of these two values.
Only the excellent frequency resolution and high accuracy of the two-year data made it possible to detect this.

Could the low frequencies simply be combination frequencies of the $\ell$\,=\,1 triplets? We have previously shown from long-term ground-based data for the $\delta$\,Sct star 44 Tau,
that the amplitude changes of the combination modes mirror the amplitude changes of their parent modes \citep{BL2008} in the sense that the product of the parent amplitudes
can be used to predict the amplitude variations of the combination frequency. We have already presented arguments that the measured high \vsini\ value implies that we see the star
almost equator-on. Consequently, photometrically the $\ell$\,=\,1, $m$\,=\,0 mode, which
has a surface node line along the equator, almost cancels out. The prograde and retrograde components, however, are fully visible. Therefore, the amplitude of the central component could be much higher than observed, raising the possibility of visible combination frequencies involving the central component.

Consequently, the low frequencies  f$_2$ and f$_3$ can be combination frequencies, rather than gravity modes. Such an interpretation is fully
supported by the related quarterly amplitude variability shown in Figure~\ref{lowamp}. Even the puzzling result of Q10 is explained, viz. the amplitude of (f$_2$+f$_3$) has increased, while the amplitudes of
both f$_2$ and f$_3$ have decreased. In the combination frequency model, f$_2$ and f$_3$ are the differences between $m$\,=\,0 and $m$\,=\,1, -1, respectively. Since the amplitude of the $m$\,=\,0
mode has decreased to nearly zero, the combinations are weak too. However, the amplitudes of $m$\,=\,1 and -1 have increased, so that the amplitude of the difference (=\,f$_2$+f$_3$) has also
increased. So the opposite behavior of f$_2$ and f$_3$ vs. (f$_2$+f$_3$) is explained.

Of course, this remarkable agreement does not prove that the low frequencies are combinations, only that they behave like combinations, usually interpreted as nonlinearities or resonant excitation in the star.

\subsection{Eclipses?}

Could the frequency regularities be caused by eclipses? In the Fourier domain, the nonsinusoidal nature of eclipse light curves
lead to long sequences of frequency multiples of the orbital frequency. In such
cases inspection of the actual light curves may be more instructive. We have examined the light curves and find no evidence for eclipses.
This does not prove that eclipses do not exist, since they could be hidden in the pulsations.

There are two further arguments against eclipses: we have already determined that the S sequence with the equidistant f$_1$ spacing does not consist of exact
multiples of f$_1$. The offset argues against a Fourier series of a nonsinusoidal light curve. Furthermore, the values of the observed low frequencies are
compatible with the size of the stellar radius, i.e., a stellar component would have to be unrealistically close to the main star.

\subsection{Spots and surface inhomogeneities}

We have already seen that the low frequencies f$_1$ to f$_4$ are within the range of the predicted rotational frequency of the star. In cooler stars, such as the Sun, spots together with
differential rotation cause light variations with similar frequencies and varying amplitudes. This is also observed for KIC\,8054146. We also note that the low frequencies are constant, i.e., if there were any spots, they would not be moving in latitude. Spots can also lead to asymmetric light curves, i.e.,
the observed frequencies between 5.5 and 6.0\,cd$^{-1}$, but fail in the details such as the peaks not being simple
harmonics of the lower frequencies. Even worse, spots cannot explain the observed close relationship
between the low and high frequencies.

\subsection{High-degree nonradial gravity modes}

Since KIC\,8054146 is a rapid rotator,
we have to take into account that the frequencies in the corotating reference system of the star may be quite different to the observed frequencies.
The transformation from the corotating system to the observer's system causes the  frequency separation between rotationally split components
to shift by an additional $mf_{\rm rot}$ factor. In case of KIC\,8054146, with $f_{\rm rot}$ close to 2.8\,cd$^{-1}$ this shifts prograde modes of  higher $m$ to rather high frequencies.

If high-degree prograde gravity modes are excited in the star, the equidistant high-frequency S sequence may in reality be a low-frequency
sequence shifted to the observer's frame of reference. In a subsequent investigation, we will model and test the work hypothesis of prograde high-degree nonradial gravity modes
(i.e., Kelvin modes) causing the S sequences.

\subsection{Asymptotic pulsation}

It is not expected that the equidistant S sequence with frequencies between 76 and 99\,cd$^{-1}$ is due to acoustic asymptotic pulsation,
because the \ion{He}{2} zone cannot drive modes at such high frequencies. Moreover, no magnetic field has been detected,
making roAp-type driving in the \ion{He}{1}/\ion{H}{0} partial ionization zone unlikely as well. Furthermore, while the distribution of mode frequencies within the asymptotic regime is predicted to be quite regular, the models do not indicate \emph{exact} equidistance.
\citet{Reese08} examined the acoustic mode frequencies in rapidly rotating stars based on polytropic models. As can be seen in Fig. 1 in their paper, even at high radial orders there is still a small variation of the frequency spacings. This deviation from exact equidistance will increase if a more realistic stellar model (including the sharp chemical discontinuity at the border of convection zones) is used. Although the departure from equidistance is small enough not to destroy the regular patterns, it should be clearly significant with the accuracy of the given data set. Consequently, the exactly equidistant frequency spacings of the S sequence is unlikely to be due to asymptotic acoustic pulsation.

\subsection{Conclusion} KIC\,8054146 was shown to be a rapidly rotating star near the cool edge of the instability strip (\Teff\,=\,7600\,$\pm$\,200\,K, \logg\,=\,3.9\,$\pm$\,0.3). The photometric {\it Kepler} data covering two years revealed a large
range of frequencies covering 200\,cd$^{-1}$. Of the 349 statistically significant detected frequencies, a large group of 117 frequencies have values less than 7\,cd$^{-1}$. Most of these are concentrated in four specific frequency bands.
They do not show the equidistant period spacings predicted for gravity modes of successive radial order, $n$, which was reported in at least one other hybrid pulsator. The four dominant low frequencies in the 2.8 to 3.0\,cd$^{-1}$  (32 to 35\,$\mu$Hz) show strong amplitude variability with a time scale of months and years.

The four dominant low frequencies also determine the spacing of the higher frequencies in and beyond the $\delta$\,Sct pressure-mode frequency domain. In fact, most of the higher frequencies belong to one of three families with spacings linked to a specific dominant low frequency. In the Fourier spectrum, these family regularities show up as triplets, high-frequency sequences with absolutely equidistant frequency spacings, side lobes (amplitude modulations) and other regularities in frequency spacings. Within each family the amplitude variations between the low and high frequencies are related. We conclude that the low frequencies (gravity modes, rotation) and observed high frequencies (mostly pressure modes) are physically connected. This unusual behavior may be related to the very rapid rotation of the star.

\acknowledgments

MB is grateful to E. L. Robinson, K. Zwintz and M. Montgomery for helpful discussions. This investigation has been supported
by the Austrian Fonds zur F\"{o}rderung der wissenschaftlichen Forschung through project P 21830-N16. 
The authors wish to thank the {\it Kepler} team for their generosity in
allowing the data to be released to the {\it Kepler} Asteroseismic Science Consortium
(KASC) ahead of public release and for their outstanding efforts which have
made these results possible.  Funding for the {\it Kepler} mission is provided
by NASA's Science Mission Directorate.

	


 
\end{document}